# DEVELOPING A 50 MeV LPA-BASED INJECTOR AT ATHENA FOR A COMPACT STORAGE RING*

E. Panofski#, C. Braun, J. Dirkwinkel, L. Hübner, T. Hülsenbusch, A. Maier, P. Messner,
J. Osterhoff, G. Palmer, T. Parikh, A. Walker, P. Winkler
Deutsches Elektronen Synchrotron, 22607 Hamburg, Germany
T. Eichner, L. Jeppe, S. Jalas, M. Kirchen, M. Schnepp, M. Trunk, C. Werle
Universität Hamburg, 20148 Hamburg, Germany
E. Bründermann, B. Härer, A.-S. Müller, C. Widmann
Karlsruhe Institute of Technology, 76131 Karlsruhe, Germany
M. C. Kaluza, A. Sävert, Helmholtz Institute Jena, 07743 Jena, Germany

## Abstract

The laser-driven generation of relativistic electron beams in plasma and their acceleration to high energies with GV/m-gradients has been successfully demonstrated. Now, it is time to focus on the application of laser-plasma accelerated (LPA) beams. The "Accelerator Technology HElmholtz iNfrAstructure" (ATHENA) of the Helmholtz Association fosters innovative particle accelerators and high-power laser technology. As part of the ATHENAe pillar several different applications driven by LPAs are to be developed, such as a compact FEL, medical imaging and the first realization of LPA-beam injection into a storage ring. The latter endeavour is conducted in close collaboration between Deutsches Elektronen-Synchrotron (DESY), Karlsruhe Institute of Technology (KIT) and Helmholtz Institute Jena (HIJ).

In the cSTART project at KIT, a compact storage ring optimized for short bunches and suitable to accept LPA-based electron bunches is in preparation. In this conference contribution we will introduce the 50 MeV LPA-based injector and give an overview about the project goals. The key parameters of the plasma injector will be presented. Finally, the current status of the project will be summarized.

## INTRODUCTION

Plasma accelerators successfully extended the portfolio of particle accelerators in the last ten years.

Beside a small footprint of the setup and a high particle energy that can be achieved, LPAs are able to generate electron beams with short bunch lengths and high peak currents. In particular this feature makes them attractive as injector for a ring based light source. Since the pulse length of the emitted photon beam is directly proportional to the bunch length of the emitting electron bunch in the ring, physics on much shorter time scales can be investigated with the radiated photon beam [1]. However, a high-quality, stable and reproducible electron beam must be generated at the laser-driven plasma injector, transported to the ring, successfully injected and, finally, stored in the synchrotron. In advance of operating such an innovative setup for user experiments, an R&D facility will be used for proof-of-principle studies and first tests. This paper introduces a 50 MeV LPA-based injector for a compact storage ring. The development of the plasma-injector is part of the ATHENA program that will be presented in the first chapter of this publication. A general discussion of the main purposes of the project follows. It is foreseen to use the large acceptance, compact storage ring of the cSTART project at KIT as a test facility for first injection studies of an LPA-based injector. This paper further describes the main parameters and special properties of the cSTART storage ring. Finally, the plans for the plasma injector prototype at DESY are presented and first simulations studies are shown, which already demonstrate the required electron beam quality.

## ATHENA PROJECT

Funded in 2018, the ATHENA program [2] provides a research and development platform for all six Helmholtz accelerator institutes in Germany working closely together with several German universities (see Fig. 1).

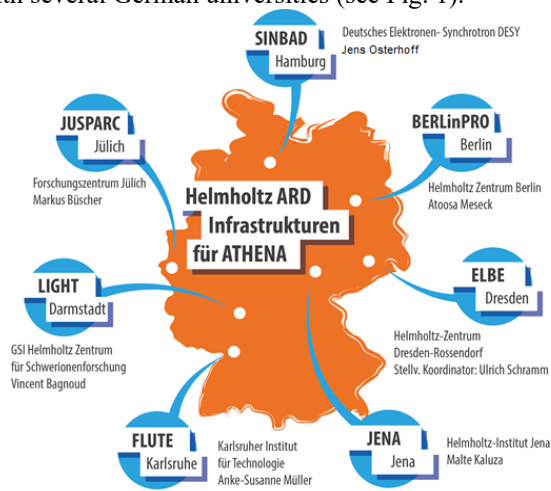

Figure 1: The ATHENA landscape in Germany with all six Helmholtz centers involved in the project.

The research in ATHENA focusses on innovative plasma-based particle accelerators and ultra-modern laser technology. In two flagship projects studies are done on electron plasma acceleration and high repetition rate laser technology at DESY Hamburg (ATHENA$_e$), while hadron plasma acceleration is investigated at Helmholtz Zentrum Dresden-Rossendorf (ATHENA$_h$). The overall goal is to demonstrate the applicability of plasma-based accelerators in different fields, ranging from a compact free electron laser, through medical uses to the injection of a plasma-produced beam into a storage ring. The latter endeavour is a joint project of DESY, HIJ and KIT and will be subject of this contribution.

___________________________________________
* Work supported by ATHENA, a project of the Helmholtz Association.
# eva.panofki@desy.de

## LASER-PLASMA-INJECTOR PROJECT

Following the mission of the ATHENA program, first studies will be performed to operate a laser-plasma accelerator as a particle injector. The final goal is to demonstrate the injection and storage of a plasma generated and accelerated electron beam into a ring-based light source. In the cooperation of DESY, HIJ and KIT, the prototype will be designed, setup and commissioned at DESY Hamburg. The clear focus will be on the generation of a high-quality beam and its characterization. After the electron beam parameters fulfill defined requirements, the facility will be used for injection into the large energy-acceptance, compact storage ring of the cSTART project at KIT.

### The cSTART Project at KIT

The compact STorage ring for Accelerator Research and Technology (cSTART) project will install a very large acceptance compact storage ring at KIT in the upcoming years [3, 4, 5]. One aim of the test facility is to handle an electron beam injected from an LPA. Since plasma-based particle accelerators usually show a larger energy spread beam with an increased divergence compared to beams accelerated by RF systems, the lattice is designed for an increased energy acceptance of $\pm$ 5% [6]. In addition, the dynamic aperture of the ring is set to 25 mm in the horizontal and 14 mm in the vertical plane in order to achieve stable storage of the LPA-beam. Furthermore, the cSTART facility is optimized for sub-ps bunches and diagnostics R&D. This feature fosters the injection of LPA-beams that typically provide bunch lengths in the fs regime. The injection energy is optimized for 50 MeV. In general, much higher energies in the range of several hundreds of MeV could be achieved with an LPA-based injector and a suitable plasma target design. However, the low energy has been chosen in order to allow the comparison of the plasma-injector with the second, conventional injector at cSTART called FLUTE [7], a 50 MeV short-pulse linear accelerator, which will be used for the beam commissioning of cSTART.

cSTART is currently in the design phase. The latest lattice design of the compact storage ring is shown in Fig. 2 [6]. Parameters of the storage ring are listed in Table 1. Studies of the longitudinal beam dynamics are ongoing [8]. A detailed design report will be published soon.

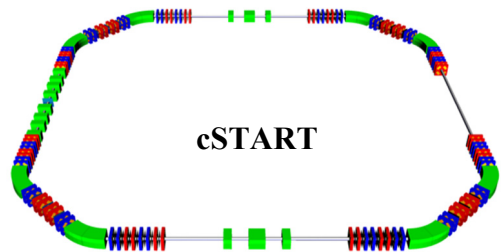

Figure 2: Lattice of the compact storage ring in the cSTART project at KIT (dipoles green, focusing/defocusing quadrupoles blue/red, sextupoles not shown). Courtesy Jens Schäfer (KIT).

Table 1: Parameters of the cSTART Lattice

| Parameters | cSTART Lattice |
| --- | --- |
| Circumference | 44.4 m |
| Beam energy | 50 MeV |
| Momentum acceptance | $\pm 5\%$ |
| Momentum compaction | $(-10$ to $+2)\cdot 10^{-3}$ |
| Dynamic aperture (horz./vert.) | 25 mm/14 mm |

### LPA Injector Prototype at DESY

A prototype of the 50 MeV plasma-injector will be built at DESY Hamburg in the upcoming months and years. The main tasks are to develop a suitable design, to set up the beamline at DESY, to generate a stable and reproducible electron beam and to measure the main beam parameters. Afterwards, the electron beam quality is optimized in a way that it fulfills the requirements for particle beam injection into the storage ring of cSTART.

The project has been started in spring 2020. The plasma-injector is currently in the design phase. Furthermore, the infrastructure is prepared for the injector installation at DESY. The setup of the beamline is planned at SINBAD [9], the former DORIS tunnel at DESY, for the year 2022. After the commissioning, the plasma-injector will be optimized and operated for injector R&D purposes. Beginning of 2025 the LPA-based injector is planned for transfer to the test facility site at KIT. For the injection into the cSTART storage ring, the beamline will be installed in the FLUTE experimental hall.

The setup of the injector prototype at DESY will include the laser focusing and the plasma target. Both will be placed in one vacuum chamber. A focusing element, such as a quadrupole duplet or an active plasma lens, will catch the electrons coming out of the target. A collimated beam is afterwards sent to a short diagnostics section, which enables to measure and optimize the electron beam quality.

The plasma-injector prototype can be operated with the KALDERA laser at DESY, which shall reach in final stages repetition rates in the kHz-range. The first operation phase of KALDERA will provide similar laser parameters, although with enhanced stability, compared to the commercial drive laser to be installed at KIT. A laser pulse energy of 500 mJ and a pulse length of 30 fs will be used for operation. The commercial laser will provide pulses up to 1.5 J at a repetition rate of up to 10 Hz. The drive laser at KIT will be upgraded at a later stage with a frontend developed at DESY to improve stability.

Table 2 displays the required electron beam parameters defined by the boundary conditions of the cSTART storage ring for a successful injection, of the laser and of the second injector FLUTE at KIT. In order to compare the performances of the short-pulse injector FLUTE and the plasma-injector, an injection energy of 50 MeV is chosen. The bunch charge of the injected beam should achieve at least 10 pC for proper diagnostics of the stored beam. The momentum acceptance of the ring defines the maximum relative energy spread of 5%. The dynamic aperture of the ring allows a normalized emittance of a few mm mrad.

Table 2: Required Electron Beam Parameters for Injection into the cSTART Storage Ring

| Parameters | cSTART Lattice |
| --- | --- |
| Energy | 50 MeV |
| Repetition rate | 10 Hz |
| Bunch charge | 10 pC |
| RMS energy spread | < 5% |
| Norm. emittance | few mm mrad |

First plasma simulations are prepared in order to check if the required electron beam parameters listed in Table 2 can be achieved.

## FIRST TARGET DESIGN AND PLASMA SIMULATIONS

As a first step during the planning of the plasma-injector prototype, the plasma target is designed and simulated. A capillary target, which is filled with a nitrogen-hydrogen gas mixture, is used. A similar model has been successfully operated at the LUX plasma accelerator at DESY [10, 11]. The electron beam generation in the target relies on localized ionization injection in combination with operating at optimal beam loading conditions [12].

The laser-plasma interaction is modelled with FBPIC, a spectral, quasi-cylindrical particle-in-cell code [13]. Additionally, the target design and the setting of the drive laser is optimized to reach the best electron beam quality by applying Bayesian optimization [14].

Figure 3 represents the best result of the simulations. The longitudinal phase space of the electron bunch is shown at the target exit. The corresponding electron beam parameters are summarized in Table 3. The simulated electron beam provides the demanded beam energy, repetition rate and bunch charge for injection into the cSTART storage ring. The transverse normalized emittance ranges in the sub-µm regime. The minimized beam divergence of 1 to 2 mrad allows to control the chromatic emittance growth between the target exit and the focusing element that collimates the beam. Although the total beam energy is low, the relative energy spread could be reduced to less than 3% (Gaussian fit) using Bayesian optimization.

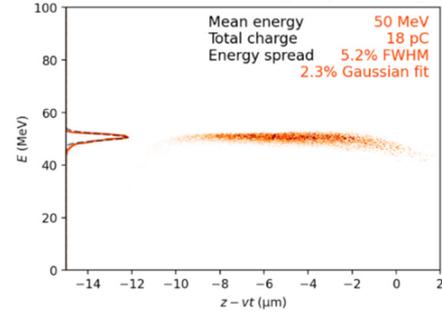

Figure 3: Longitudinal phase space of the FBPIC-simulated electron beam at the target exit.

Table 3: Simulation Results of the Current Target Design. Electron Beam Parameters are Displayed at the Target Exit

| Parameters | Simulation Results |
| --- | --- |
| Energy | 50 MeV |
| Repetition Rate | 10 Hz |
| Bunch charge | (20 ± 5) pC |
| Rel. energy spread (Gaussian fit) | 2 to 3% |
| Norm. emittance | sub-µm range |
| RMS divergence | 1 to 2 mrad |

The promising results of the first laser-plasma simulations indicate that the plasma-injector is able to deliver a high-quality beam suitable for the injection into cSTART. Furthermore, a compact, low energy-spread beam will allow a controlled transport of the electrons from the injector to the injection point. Start-to-end simulations are planned to be done at KIT. The first plasma target for the injector commissioning will be manufactured in the next weeks. A hexapod, which has been already designed, will allow the target alignment relative to the drive laser during the experiment. As a next step, the design of the vacuum chamber, which houses the parabola for laser focusing, pre-target laser diagnostics and the plasma target with the alignment system, is going to be finished.

## SUMMARY

This contribution introduces a 50 MeV LPA-based injector for the injection into the large acceptance, compact storage ring of the cSTART project at KIT. The prototype of the plasma injector is currently in the design phase and will be installed and operated at DESY in the upcoming months and years. Simulations with FBPIC have been successfully demonstrated the target electron beam parameters that are required for electron beam injection into cSTART. A fully operational and optimized laser-driven plasma injector is expected at KIT beginning of the year 2025.